\documentclass[12pt]{article}
\usepackage{amssymb,amsmath}
\usepackage[noblocks]{authblk}
\usepackage[top=0.75in, bottom=0.75in, left=0.75in, right=0.75in, dvips]{geometry}
\usepackage{caption}
\begin{document}
\textwidth 10.0in 
\textheight 9.0in 
\topmargin -0.60in
\title{Derivation of All Symmetries from First-Class Constraints and Quantization in $2+1$ Dimensional Supergravity}
\author[1,2]{D.G.C. McKeon}
\affil[1] {Department of Applied Mathematics, The
University of Western Ontario, London, ON N6A 5B7, Canada} 
\affil[2] {Department of Mathematics and
Computer Science, Algoma University, Sault St.Marie, ON P6A
2G4, Canada}
\date{}
\maketitle

\maketitle
\noindent
email: dgmckeo2@uwo.ca

\begin{abstract}
The first class constraints in $N = 1$ supergravity in $2 + 1$ dimensions are used to construct a generator of three gauge symmetries (including a local supersymmetry) that leave the action invariant.  The algebra of these symmetries closes.  This generator is used to quantize the model; a ghost involving both Bosonic and Fermionic components arises.
\end{abstract}

\section{Introduction}
Gravity in $2 + 1$ dimensions has been extensively treated [1-3] and its supersymmetric extension considered [4].  In these discussions, the local symmetries of the theory have been treated as being manifest.  However, it has long been understood that local symmetries (or ``gauge'' symmetries) are closely linked with the presence of first-class constraints that arise when using the Dirac constraint formalism to analyze their canonical structure [5].

Two approaches have been used to derive these gauge symmetries from the first-class constraints.  In the ``HTZ'' approach [6], symmetries in the Hamiltonian form of the action are examined directly while in the ``C'' approach [7] symmetries present in the equations of motion are considered.  Normally, only Bosonic symmetries have been determined in either of these approaches, but recently Fermionic (or ``super'') symmetries have been shown to follow from the presences of Fermionic first-class constraints [8].  The spinning particle was used to demonstrate this; we now will employ the Fermionic and Bosonic first-class constraints present in $2 + 1$ dimensional supergravity to find one Fermionic and two Bosonic symmetries present in the model.  The algebra of these gauge symmetries will be shown to close. Quantization is effected through use of the path integral, taking into account both the first-and second-class constraints present.

Conventions used are given in the appendix.

\section{Supergravity in $2 + 1 D$}

We work with the first order Lagrangian 
\begin{equation}
\mathcal{L}= \epsilon^{\mu\nu\lambda} \left( b_\mu^i R_{\nu\lambda  i} + \overline{\psi}_\mu D_\nu \psi_\lambda\right).
\end{equation}
There are independent Boson fields $(b_\mu^i , w_\mu^i)$ and the Fermion field $\psi_\mu$ with 
\begin{equation}
R_{\mu\nu i} = \partial_\mu w_{\nu i} - \partial_\nu w_{\mu i} - \epsilon_{ijk} w_\mu^j w_\nu^k 
\end{equation}
and 
\begin{equation}
D_\mu = \partial_\mu + \frac{i}{2} \gamma^i w_{\mu i}
\end{equation}
so that $[D_\mu , D_\nu] = \frac{i}{2} \gamma^i R_{\mu\nu i}$.

From eq. (1) it follows that the momenta conjugate to $(b^{0i} \equiv b^i, b^{\alpha i}, w^{0i} \equiv w^i, w^{\alpha i}, \psi_0 \equiv \psi, \psi_\alpha)$ are respectively 
\begin{subequations}
\begin{align}
p_i &= 0 \\
p_{\alpha i} &= 0 \\
I\!\!P_i &= 0 \\
I\!\!P_{\alpha i} &= 2\epsilon_{\alpha\beta} b^\beta_{\;\,i} \\
\pi &= 0 \\
\pi^\alpha &= -\epsilon^{\alpha\beta}\overline{\psi}_\beta .
\end{align}
\end{subequations}
The constraints of eqs. (4b,d) and (4f) are obviously second class; they result in the Dirac bracket (DB)
\begin{align}
\left\lbrace A,B \right\rbrace^* = \left\lbrace A,B \right\rbrace + \frac{1}{2} \epsilon_{\alpha\beta} &  \bigg[ \left\lbrace A, \pi^\alpha + \epsilon^{\alpha\gamma} \overline{\psi}_\gamma \right\rbrace \gamma^0 
\left\lbrace \left( \pi^\beta + \epsilon^{\beta \delta} \overline{\psi}_\delta \right)^T, B\right\rbrace \nonumber \\
& + \left\lbrace A, I\!\!P^{\alpha i} - 2 \epsilon^{\alpha\gamma} b_\gamma^{\,i}\right\rbrace \left\lbrace p^\beta_{\;i}, B\right\rbrace \nonumber \\
& - \left\lbrace A, p^{\alpha i} \right\rbrace \left\lbrace I\!\!P^{\beta} _{\;i} - 2 \epsilon^{\beta\gamma} b_{\gamma i}, B \right\rbrace \bigg]
\end{align}
where $A$ and $B$ are dynamical variables.  From eq. (5) it follows that 
\begin{subequations}
\begin{align}
\left\lbrace b_\alpha^i, w_\beta^j \right\rbrace^* &= \frac{1}{2} \eta^{ij} \epsilon_{\alpha\beta} \\
\left\lbrace \psi_\alpha, \overline{\psi}_\beta \right\rbrace^* &= \frac{1}{2}  \epsilon_{\alpha\beta}.
\end{align}
\end{subequations}
The canonical Hamiltonian now is given by
\begin{align}
\mathcal{H}_c = \epsilon^{\alpha\beta} \bigg[ -b_i R_{\alpha\beta}^i &- 2w_i \left(\partial_\alpha b_\beta^i - \epsilon^{ijk} w_{\alpha j} b_{\beta k} - \frac{i}{4} \overline{\psi}_\alpha \gamma^i \psi_\beta \right) \nonumber \\
&- 2 \overline{\psi} (D_\alpha \psi_\beta ) \bigg].
\end{align}
With this $\mathcal{H}_c$, the primary constraints of eqs. (4a,c,e) lead to the secondary constraints
\begin{subequations}
\begin{align}
\Phi_1^i &= \epsilon^{\alpha\beta} R_{\alpha\beta}^i \\
\Phi_2^i &= \epsilon^{\alpha\beta} \left( \partial_\alpha b_\beta^i - \epsilon^{ijk} w_{\alpha j} b_{\beta k} - \frac{i}{4} \overline{\psi}_\alpha \gamma^i \psi_\beta \right)\\
\intertext{and} 
\Psi & = \epsilon^{\alpha\beta} D_\alpha\psi_\beta 
\end{align}
\end{subequations}
respectively.  These primary and secondary constraints are all first class and there are no higher generation constraints because of the algebra
\begin{subequations}
\begin{align}
\left\lbrace \Phi_1^i, \Phi_2^j \right\rbrace^* &= -\frac{1}{2} \epsilon^{ijk} \Phi_{1k} \\
\left\lbrace \Phi_2^i, \Phi_2^j \right\rbrace^* &= -\frac{1}{2} \epsilon^{ijk} \Phi_{2k} \\
\left\lbrace \Psi, \overline{\Psi} \right\rbrace^* &= -\frac{i}{8}  \Phi_{1i} \gamma^i \\
\left\lbrace \Psi, \Phi_2^i \right\rbrace^* &= \frac{i}{4}   \gamma^i \Psi
\end{align}
\end{subequations}
with all other DB between constraints vanishing.

It is possible to show that the HTZ approach [6] to finding gauge symmetries in a theory from first-class constraints can be used even when Fermionic constraints are present. (In this case, these are the primary constraint $\pi = 0$ and the secondary constraint $\Psi = 0$.)  The form of the generator of gauge symmetries is 
\begin{equation}
G = \int d^2x \left[ A_1^i p_i + A_2^i \Phi_{1i} + B_1^i I\!\!P_i + B_2^i \Phi_{2i} + C_1^T\pi^T + \overline{C}_2 \Psi \right] 
\end{equation}
where $(A_1^i, B_1^i, A_2^i, B_2^i)$ are Bosonic and $(C_1, C_2)$ are Fermionic.  It follows from the DB
\begin{align}
\left\lbrace G, \int d^2y \,\mathcal{H}_c \right\rbrace^* = &\int d^2x \bigg[ \left( A_{1k} + \epsilon_{ijk} \left( A_2^i w^j + \frac{1}{2} B_{2i} b^j\right) - \frac{i}{4} \overline{C}_2 \gamma_k \psi \right)\Phi_1^k \nonumber \\
&+ \left(  2 B_{1k} + \epsilon_{ijk} B_2^i w^j\right)\Phi_2^k \nonumber \\
&+ \left( 2 \overline{C}_1 + \frac{i}{2} \left(B_2^i \overline{\psi} \gamma_i - w_i \overline{C}_2 \gamma^k   \right) \right) \Psi \bigg] 
\end{align}
that the HTZ equation leads to 
\begin{subequations}
\begin{align}
& \dot{A}_{2k} + A_{1k} + \epsilon_{k\ell m} (A_2^\ell w^m + \frac{1}{2} B_2^\ell b^m) + \frac{i}{4} \overline{C}_2 \gamma_k \psi = 0 \\
& \dot{B}_2^k + 2B_{1k} + \epsilon_{k\ell m} B_2^\ell w^m = 0 \\
& \dot{\overline{C}}_2 + 2\overline{C}_1 + \frac{i}{2} \left( \overline{\psi} B_2\cdot \gamma - \overline{C}_2 w \cdot \gamma\right) = 0.
\end{align}
\end{subequations}
From eq. (12), we find now that $G$ is given by 
\begin{align}
G = \int d^2x & \bigg[ - \left( \dot{A}_k + \epsilon_{k\ell m} (A^\ell w^m + \frac{1}{2} B^\ell b^m) + \frac{i}{4} \overline{C}\, \gamma_k\psi \right)p^k \nonumber \\
& - \frac{1}{2} \left( \dot{B}_k + \epsilon_{k\ell m} B^l w^m \right)I\!\!P^k \nonumber \\
& - \frac{1}{2} \left( \dot{\overline{C}} + \frac{i}{2} (\overline{\psi} B \cdot \gamma - \overline{C} w \cdot \gamma)  \right)\gamma^0 \pi^T \nonumber \\
& \hspace{.5cm} + A_k \Phi_1^k + B_k \Phi_2^k + \overline{C} \Psi\bigg].
\end{align}

We now will establish the DB algebra of the generator $G$; that is, if $G_I$ is associated with gauge functions $(A_I, B_I, C_I)$ we wish to compute $\left\lbrace G_I, G_J \right\rbrace^*$.  This can be done directly, but it is easier to make use of the following general argument.  If one has a set of canonical variables $(Q_i, I\!\!P_i)$, $(q_i, p_i)$ after making use of the second class constraints, and if the canonical Hamiltonian is of the form 
\begin{equation}
H_c = - Q_i \Phi_i (q,p)
\end{equation}
with
\begin{equation}
\left\lbrace \Phi_i (q,p), \Phi_j (q,p)\right\rbrace^* = c_{ijk} \Phi_k (q,p)
\end{equation}
then the gauge generators associated with the first class constraints $(I\!\!P_i, \Phi_i(q,p))$ is by the HTZ approach [6]
\begin{equation}
G_I = \left( -\dot{\Lambda}_{Ii} + c_{ijk} \Lambda_{Ij} Q_k  \right) I\!\!P_i + \Lambda_{Ii} \Phi_i (q,p)
\end{equation}
where $\Lambda_{Ii}(t)$ is the gauge parameter.  It may be shown now that upon using the Jacobi identity for the quantities $c_{ijk}$ that follow from eq. (15) that 
\begin{equation}
\left\lbrace G_I, G_J \right\rbrace^* = G_K
\end{equation}
where
\begin{equation}
\Lambda_{Ki} = c_{ijk} \Lambda_{Ij} \Lambda_{Jk}.
\end{equation}
The model based on eq. (1) is consistent with eqs. (14, 15); we then find that eq. (18) shows that with $G$ given by eq. (13), then $G_K$ in eq. (17) is given by 
\begin{subequations}
\begin{align}
A_{Ki} &= -\frac{i}{8} \overline{C}_I \gamma_i C_J + \frac{1}{2} \epsilon_{ijk} \left( A_J^j B_I^k - A_I^j B_J^k\right)\\
B_{Ki} &= -\frac{1}{2} \epsilon_{ijk} B_I^j B_J^k \\
\overline{C}_{K} &= \frac{i}{4} \left( \overline{C}_I B_J \cdot \gamma - \overline{C}_J B_I \cdot \gamma \right)
\end{align}
\end{subequations}
upon using the structure functions $c_{ijk}$ that follow from eq. (9). The gauge algebra closes without having to introduce auxiliary fields.

It follows from eq. (13) that the variations in $\psi_\mu$, $b_\mu^i$ and $w_\mu^i$ that are generated by $G$ are
\begin{subequations}
\begin{align}
\delta\psi_\mu &= -\frac{1}{2} \left( D_\mu C - \frac{i}{2} B \cdot \gamma \psi_\mu \right) \\
\delta b_\mu^i &= -\left[ \mathcal{D}_\mu^{ij} A_j -  \frac{1}{2} \epsilon^{ijk} b_{j\mu} B_k + \frac{i}{4} \overline{C}  \gamma^i \psi_\mu\right] \\
\intertext{and}
\delta w_\mu^i &= -\frac{1}{2}\mathcal{D}_\mu^{ij} B_j\;, 
\end{align}
\end{subequations}
where 
\begin{equation}
\mathcal{D}_\mu^{ij} = \partial_\mu \delta^{ij} - \epsilon^{ipj} w_{p\mu} \left(\left[ \mathcal{D}_\mu , \mathcal{D}_\nu \right]^{ij} =\epsilon^{ijp} R_{\mu\nu p}\right).
\end{equation}

We now turn to the problem of quantizing this model using the path integral.

\section{Quantization}

In a model with variables $(q_i(t), p_i(t))$ in phase space, and governed by a canonical Hamiltonian $H_c(q_i, p_i)$, quantization is effected through the path integral for the transitional amplitude
\begin{equation}
<\text{out}|\text{in}> = \int Dp_i\, Dq_i \,\exp i \int_{-\infty}^\infty dt (\dot{q}_i p_i - H_c) 
\end{equation}
where $q_i(t) \rightarrow (q_i^{\text{out}}, q_i^{\text{in}})$ as $t \rightarrow \pm \infty$ [9].  In the presence of first-class [16] and second-class constraints [17], the measure for this path integral receives the contribution
\begin{equation}
M = \det \left\lbrace \phi_i, \gamma_j \right\rbrace \mathrm{det}^{1/2} \left\lbrace \theta_i, \theta_j \right\rbrace \delta (\phi_i) \delta(\gamma_i) \delta(\theta_i) 
\end{equation}
where $\phi_i$ is the set of first-class constraints, $\gamma_i$ are the associated gauge conditions and $\theta_i$ are the set of second-class constraints.

With the second class constraints of eqs. (4b,d,f) we see that $\left\lbrace\theta_i, \theta_j \right\rbrace$ is field independent and so in eq. (23) the second-class constraints just serve to rescale $M$ by a constant.  (This is unlike the model discussed in ref. [10].)

The first class constraints which lead to the contributions $\det \left\lbrace\phi_i, \gamma_j\right\rbrace\delta(\phi_i)\delta(\gamma_i)$ in eq. (23) can be handled in an alternate manner that accommodates covariant gauge fixing [11].  In this alternate approach, the Faddeev-Popov quantization procedure is adapted so as to be applicable to a path integral in phase space.

The constant factor
\begin{align}
K = \int DA_i\, & DB_i \,DC\, \delta\left( (\mathcal{D} \cdot b)^i + \left\lbrace (\mathcal{D} \cdot b)^i, G \right\rbrace^* - k_b^i \right)\\
& \delta\left( \partial \cdot w^i + \left\lbrace \partial \cdot w^i, G\right\rbrace^* - k_w^i\right)\nonumber\\
& \hspace{1.5cm} \delta \left( (D \cdot \psi ) + \left\lbrace (D \cdot \psi), G \right\rbrace^* - k_\psi \right)\Delta \nonumber
\end{align}
is first defined.  From the changes in $\psi_\mu$, $b_\mu^i$ and $w_\mu^i$ induced by $G$ that are given in eq. (20), it follows that 
\begin{equation}
\Delta = s\,\det \left(
\begin{array}{ccc}
(\mathcal{D}^2)^{ij} & -\frac{1}{2}(\mathcal{D}_\mu^{ip})(\epsilon_{pq}^{\;\;\;\;\,\ell} b^{q\mu}) & -\frac{i}{4} \mathcal{D}_\mu^{ip} \overline{\psi}^\mu \gamma_p \\
 & &\\
0 & \frac{1}{2}\partial \cdot \mathcal{D}^{k\ell} & 0 \\
& & \\
0 & - \frac{i}{2} D^\mu\gamma^\ell\psi_\mu & \frac{1}{2} D^2 
\end{array}
\right)
\end{equation}
in order that $K$ be a constant.

We now introduce $K$ into the phase-space path integral whose form is given by eq. (22) with the canonical Hamiltonian of eq. (7). The change of variables induced by $-G$ is then performed; this leaves the action in phase space unaltered.  Upon introducing the integrals over the field independent quantities $k_b^i$, $k_w^i$ and $k_\psi$ as in ref. [12]
\begin{equation}
\overline{K} = \int Dk_b^i\,Dk_w^i \, Dk_\psi \exp - \frac{1}{2} \int dx \left[ (k_b^i)^2 + (k_w^i)^2 + \overline{k}_\psi k_\psi \right]
\end{equation}
and converting the phase space path integral to a configuration space path integral using the approach of ref. [13], we are left with the transition amplitude 
\begin{align}
<\text{out}|\text{in}> = \int Db_\mu^i\, Dw_\mu^i\,&  D\psi_\mu  \exp i \int dx \bigg[\mathcal{L} - \frac{1}{2}(\partial \cdot w^i)^2 \\
& - \frac{1}{2} (\mathcal{D}^{ij} \cdot b_j)^2 - \frac{1}{2} (\overline{D \cdot \psi})(D \cdot \psi) \bigg] \Delta \nonumber
\end{align}
with $\mathcal{L}$ being given by eq. (1).

The functional determinant of eq. (26) can now be exponentiated using Fermionic ghost fields $(\overline{c}_i, d_j)$ and $(\overline{d}_i, d_j)$ and the Bosonic ghost field $\Gamma$.  (The Fermionic ghost fields are all vectors and the Bosonic ghost field is a Dirac spinor in the tangent space.)  We have 
\begin{equation}
\Delta = \int D\overline{c}_i\, Dc_j\,  D\overline{d}_i\, Dd_j \,D\Gamma
\exp i \int dx \left( \overline{c}_i,\, \overline{d}_k, \,\overline{\Gamma} \right) \mathbf{M} (c_j, d_\ell,  \Gamma)^T
\end{equation}
where $\mathbf{M}$ is the supermatrix appearing in eq. (26).  The presence of a Fermionic gauge invariance has generalized the functional Faddeev-Popov determinant to being a superdeterminant.

\section*{Discussion}

We have examined the canonical structure of $N = 1$ supergravity in $2 + 1$ dimensions, and from the first class constraints that occur, deduced the gauge symmetries that reside in the theory.  The model is quantized using the phase space form of the path integral, in conjunction with a means of employing a covariant gauge fixing technique while working in phase space.

The form of the path integral appearing in eqs. (27, 28) could have been derived directly by applying the Faddeev-Popov approach to the configuration space form of the path integral.  However, the transition from the phase space form of the path integral (which follows from canonical quantization [9]) to the configuration space form of the path integral is not always so straight forward as it is here and consequently the Faddeev-Popov quantization procedure is not always viable.  This is the case if the model were to have second class constraints with field dependent Poisson Brackets with each other, such as the model of ref. [10] or the first order Einstein-Hilbert action in $D > 2$ dimensions [14].

If one were to compute loop corrections to the effective action in eqs. (28, 29), operator regularization as employed with Chern-Simons theory should be a convenient technique that could be used [15].

The derivation of the gauge generator from the first class constraints in the theory should provide a useful means of uncovering all gauge symmetries in higher dimensional supergravity theories, such as $N = 8$ supergravity in $D = 4$ dimensions. In this model, apparently fortuitous cancellation of divergences in higher loop calculations have been attributed to the presence of symmetries that are not manifest.

\section*{Acknowledgements}
R. Macleod had a helpful comment.

\section*{Appendix-Conventions}
We use the metric $\mathrm{diag}\; \eta^{ij} = (+ - -)$ with $\epsilon^{012} = +1$.  Dirac matrices $\gamma^i$ are related to Pauli spin matrices by $\gamma^0 = \sigma_2,$ $\gamma^1 = i\sigma_3 \, \gamma^2 = i\sigma_1$ so that
\[ \gamma^i \gamma^j = \eta^{ij} + i\epsilon^{ijk} \gamma_k \eqno(A.1) \]
and
\[ \gamma^0 \gamma^i\gamma^0 = -\gamma^{iT} = \gamma^{i\dagger}.\eqno(A.2) \]
Latin indices $(i,j,k \ldots )$ are used for the target space, Greek indices $(\mu , \nu , \lambda \ldots)$ are used for space-time indices $(0, 1, 2)$ while early Greek indices $(\alpha, \beta, \gamma \ldots)$ are used for space indices $(1,2)$.  We take $\epsilon^{12} = \epsilon^{012} = \epsilon_{12} = 1$.

All spinors $\psi$ are taken to satisfy the Majorana condition $\psi = C\overline{\psi}^T$ where $\overline{\psi} = \psi^\dagger\gamma^0$ and $C = -\gamma^0$ so that $\psi = \psi^*$ is real.  We use the equations 
\[ \overline{\chi}\phi = \overline{\phi}\chi,\quad \overline{\chi}\gamma^i\phi = - \overline{\phi}\gamma^i\chi\,. \eqno(A.3a,b) \]

For Grassmann variable $\theta_a$, we use the left derivative so that
\[ \frac{d}{d\theta_a} (\theta_b\theta_c) = \delta_{ab} \theta_c - \delta_{ac} \theta_b; \quad \frac{d}{dt} F(\theta(t)) = \dot{\theta}(t) F^\prime (\theta(t)). \eqno(A.4a,b) \]

For Bosonic fields $B_1$ and Fermionic fields $F_i$, we define Poisson brackets with respect to Bosonic canonical pairs $(q_i, p_i = \frac{\partial L}{\partial \dot{q}_i})$ and Fermionic canonical pairs
$(\psi_i,\pi_i = \frac{\partial L}{\partial \dot{\psi}_i})$ by 
\[ \left\lbrace B_1, B_2 \right\rbrace = 
\left(  B_{1,q} B_{2,p} - B_{2,q} B_{1,p}\right) + 
\left(  B_{1,\psi} B_{2,\pi} - B_{2,\psi} B_{1,\pi}\right) = - \left\lbrace B_2, B_1 \right\rbrace \eqno(A.5a)\]
\[ \left\lbrace B, F \right\rbrace = 
\left(  B_{,q} F_{,p} - F_{,q} B_{,p}\right) +
\left(  B_{,\psi} F_{,\pi} + F_{,\psi} B_{,\pi}\right) = - \left\lbrace F, B \right\rbrace \eqno(A.5b)\]
\[ \left\lbrace F, B \right\rbrace = 
\left(  F_{q,} B_{,p} - B_{,q} F_{,p}\right) -
\left(  B_{,\psi} F_{,\pi} + F_{,\pi} B_{,\pi}\right) = - \left\lbrace B, F \right\rbrace \eqno(A.5c)\]
\[ \left\lbrace F_1, F_2 \right\rbrace = 
\left(  F_{1,q} F_{2,p} + F_{2,q} F_{1,p}\right) -
\left(  F_{1,\psi} F_{2,\pi} + F_{2,\psi} F_{1,\pi}\right) = \left\lbrace F_2, F_1\right\rbrace \eqno(A.5d)\]
where $B_{,q} F_{,p} = \displaystyle{\sum_i} \frac{\partial B}{\partial q_i} \,\frac{\partial F}{\partial p_i}$ etc. It follows that 
\[ \left\lbrace XY, Z \right\rbrace = X\left\lbrace Y, Z \right\rbrace + (-1)^{\epsilon{_y}\epsilon{_z}} \left\lbrace X, Z \right\rbrace Y\eqno(A.6a) \]
\[ \left\lbrace X,YZ \right\rbrace = (-1)^{\epsilon{_x}\epsilon{_y}}
Y\left\lbrace X, Z \right\rbrace +  \left\lbrace X, Y \right\rbrace Z\eqno(A.6b) \]
where $\epsilon_x = 1$ if $X$ is Fermionic and $\epsilon_x = 0$ if $X$ is Bosonic.

The Hamiltonian is given by 
\[ H(q_i, p_i, \psi_i ,\pi_i) = \dot{q}_i p_i + \dot{\psi}_i \pi_i - L (q_i, \dot{q}_i, \psi_i, \dot{\psi}_i) \eqno(A.7) \]

\end{document}